\documentclass[a4paper]{jpconf}
\pdfoutput=1

\usepackage{citesort}
\usepackage{graphicx}
\usepackage{bm}

\newcommand{\beq}{\begin{equation}}
\newcommand{\eeq}{\end{equation}}
\newcommand{\bqa}{\begin{eqnarray}}
\newcommand{\eqa}{\end{eqnarray}}

\def\simge{\mathrel{
    \rlap{\raise 0.511ex \hbox{$>$}}{\lower 0.511ex \hbox{$\sim$}}}}
\def\simle{\mathrel{
    \rlap{\raise 0.511ex \hbox{$<$}}{\lower 0.511ex \hbox{$\sim$}}}}
    
\begin{document}


\title{Bottomonia in the Quark Gluon Plasma}

\author{Michael Strickland}

\address{Physics Department, Gettysburg College, Gettysburg, PA 17325 United States}

\begin{abstract}
I review recent calculations of the suppression of bottomonium states in heavy ion 
collisions.  A non-relativistic potential is used which is
complex valued.  This allows one to extract the binding energies and decay widths of 
the ground and excited states of bottomonium as a function of the typical plasma particle
momentum and momentum-space anisotropy.  The decay widths determined  
are used as input and integrated over space-time taking into account the dynamical evolution
of the typical particle momentum and momentum-space anisotropy.  The suppression of $\Upsilon(1s)$, 
$\Upsilon(2s)$, $\Upsilon(3s)$, $\chi_{b1}$, and $\chi_{b2}$ is obtained as a function of 
centrality, rapidity, and transverse momentum.  The obtained results are compared with data
from the STAR and CMS collaborations.
\end{abstract}


\section{Introduction}

In November of 2010 the Large Hadron Collider (LHC) at the European Organization for Nuclear 
Research (CERN) achieved its first relativistic heavy ion collisions of lead nuclei (Pb-Pb).  The goal 
of such experiments is to recreate conditions which only existed in the early universe before the 
formation of hadrons.  The LHC is searching for evidence of the creation of a primordial 
state of matter called the quark gluon plasma (QGP) which comprised the entire universe until 
approximately $10^{-5}$ seconds after the big bang when the temperature of the universe was 
on the order of $10^{12}$ Kelvin.  The LHC experiment is attempting to recreate such temperatures
by colliding lead nuclei at $\sqrt{s_{NN}} = 2.76$ TeV/nucleon whereas the lower energy 
Relativistic Heavy Ion Collider (RHIC) at Brookhaven National Laboratory has been colliding gold 
nuclei at $\sqrt{s_{NN}} = 200$ GeV/nucleon.  In addition to simply crossing the threshold for 
the creation of a quark gluon plasma, the experiments are also making detailed measurements 
of the collision products in order to determine fundamental properties of the QGP.  The 
resulting data are being compared to theoretical calculations based on quantum chromodynamics (QCD) 
which describes the interactions of quarks and gluons.

At high temperatures one expects the emergence
of Debye screening of the interaction between quarks and gluons.  This
leads to the dissolution of hadronic bound states
\cite{Shuryak:1980tp}.  A particularly interesting subset of hadronic
states consists of those comprised of heavy quarks, since the spectrum
of low lying states can be found using potential-based
non-relativistic treatments.  Based on such potential models there
were early predictions \cite{Matsui:1986dk,Karsch:1987pv} that
$J/\psi$ production would be suppressed in heavy ion collisions
relative to the corresponding production in proton-proton collisions
scaled by the number of nucleons participating in the collision.  In
recent years there have been important theoretical advances in the
understanding of heavy quark states at finite temperature using
analytic techniques
\cite{Grandchamp:2005yw,Laine:2006ns,Laine:2007gj,Dumitru:2007hy,Rapp:2008tf,Brambilla:2008cx,Dumitru:2009ni,Burnier:2009yu,Margotta:2011ta}
and lattice QCD
\cite{Umeda:2002vr,Asakawa:2003re,Datta:2003ww,Aarts:2007pk,Hatsuda:2006zz,Jakovac:2006sf,Aarts:2010ek,Aarts:2012ka,Ding:2012pt}. 
Most recently,
interest has shifted to bound states of bottom and anti-bottom quarks (bottomonium) for the following 
reasons
\begin{enumerate} 
\item
Bottom quarks ($m_b \simeq 4.2$~GeV) are more massive than charm quarks 
($m_c \simeq 1.3$~GeV) and as a result the heavy quark effective theories underpinning 
phenomenological applications are on much surer footing. 
\item
Due to their higher mass, the effects of initial state nuclear suppression are expected to
be smaller than for the charmonium states \cite{Rakotozafindrabe:2012ss}.
\item
The masses of bottomonium states ($m_\Upsilon \approx$~10 GeV) are much higher than the 
temperatures ($T \simle 1$ GeV) generated in relativistic heavy ion collisions.  As a result,  
bottomonium production will be dominated by initial hard scatterings.
\item
Since bottom quarks and anti-quarks are relatively rare within the plasma, the probability for 
regeneration of bottomonium states through recombination is much smaller than for charm 
quarks.
\end{enumerate}
As a result one expects the bottomonium system to be a cleaner probe of the quark gluon
plasma than the charmonium system for which the modeling has necessarily become quite
involved.  For this reason we will focus on the bottomonium states and only consider
the thermal suppression of these states, ignoring initial state effects and any possible thermal
generation or recombination.
In this conference proceedings we will review recent theoretical calculations of bottomonium suppression at 
energies probed in relativistic heavy ion collisions at RHIC and LHC.  We will
present an overview of the important aspects of the calculation and refer the reader to
Refs.~\cite{Strickland:2011mw,Strickland:2011aa} for details.  We will compare our predictions
for inclusive $\Upsilon(1s)$ and $\Upsilon(2s)$ suppression with recent data to from the CMS and 
STAR collaborations.  

\section{Theoretical methods}

In the last few years there have been important theoretical developments in the theoretical 
treatment of heavy quarkonium in the quark gluon plasma.  
First among these are the first-principles calculations of imaginary-valued contributions
to the heavy quark potential.  The first calculation of the leading-order perturbative imaginary 
part of the potential due to gluonic Landau damping was performed by Laine 
et al.~\cite{Laine:2006ns}.  Subsequently, an additional imaginary-valued 
contribution to the potential coming from singlet to octet transitions has also been computed 
using the effective field theory approach~\cite{Brambilla:2008cx}.  These imaginary-valued
contributions to the potential are related to quarkonium decay processes in the plasma.  The 
consequences of such imaginary parts on heavy quarkonium spectral functions 
\cite{Burnier:2007qm,Miao:2010tk}, perturbative thermal widths \cite{Laine:2006ns,%
Brambilla:2010vq}, quarkonia at finite velocity \cite{Escobedo:2011ie},
in a T-matrix approach 
\cite{Grandchamp:2005yw,Rapp:2008tf,Riek:2010py,Emerick:2011xu,Zhao:2011cv}, and
in stochastic real-time dynamics \cite{Akamatsu:2011se} have recently been studied.  

Additionally, there have been significant advances in the dynamical models used to simulate
plasma evolution.  In particular, there has been a concerted effort to understand the effects
of plasma momentum-space anisotropies generated by the rapid longitudinal expansion of
the matter along the beamline direction.  The resulting dynamical models are now able to describe the
anisotropic hydrodynamical evolution using full (3+1)-dimensional simulations 
\cite{Florkowski:2010cf,Martinez:2010sc,Ryblewski:2010bs,Martinez:2010sd,%
Martinez:2012tu,Ryblewski:2012rr}.  
This is important because 
momentum-space anisotropies can have a significant impact on quarkonium suppression
since in regions of high momentum-space anisotropy one expects reduced quarkonium binding
\cite{Dumitru:2007hy,Dumitru:2009ni,Burnier:2009yu,Dumitru:2009fy,%
Margotta:2011ta}.  In Refs.~\cite{Strickland:2011mw,Strickland:2011aa} the dynamical
evolution of the anisotropic plasma was combined with the real and imaginary parts of the binding
energy obtained using modern complex-valued potentials.

\subsection{Anisotropic potential model and binding energies}

Early on it was shown that ideal relativistic hydrodynamics is
able to reproduce the soft collective flow of the matter and single particle spectra produced at RHIC 
\cite{Huovinen:2001cy,Hirano:2002ds,Tannenbaum:2006ch,Kolb:2003dz}.  
Based on this, there was a concerted effort to develop a systematic framework for describing the soft 
collective motion.  This effort resulted in a number of works dedicated to 
the development and application of relativistic viscous hydrodynamics 
to relativistic heavy ion collisions
\cite{Muronga:2001zk,Muronga:2003ta,Muronga:2004sf,Baier:2006um,Romatschke:2007mq,Baier:2007ix,%
Dusling:2007gi,Luzum:2008cw,Song:2008hj,El:2009vj,Denicol:2010tr,Denicol:2010xn,%
Schenke:2010rr,Schenke:2011tv,Bozek:2011wa,Niemi:2011ix,Niemi:2012ry,Bozek:2012qs,Denicol:2012cn}.  

One of the weakness of the traditional viscous hydrodynamics approach 
is that it relies on an implicit assumption that the system 
is close to thermal equilibrium which implies that the system is also very close to being isotropic in momentum space. 
However, one finds during the application of these methods that this 
assumption breaks down at the earliest times after the initial impact of the two nuclei due to large 
momentum-space anisotropies in the $p_T$-$p_L$ plane which can
persist for many fm/c \cite{Martinez:2009mf}.  In addition, one finds that near the transverse and longitudinal edges of 
the system these momentum-space 
anisotropies are large at all times \cite{Martinez:2009mf,Martinez:2010sd,Ryblewski:2010bs}.  
Similar conclusions have been obtained in the context of strongly
coupled systems where it has been shown using the conjectured AdS/CFT correspondence one achieves viscous 
hydrodynamical behavior at times when the system
still possesses large momentum-space anisotropies and that these anisotropies remain large throughout the evolution 
\cite{Chesler:2008hg,Chesler:2009cy,Heller:2011ju,Heller:2012je,Heller:2012km,Wu:2011yd,Chesler:2011ds}.  
Based on these results one is motivated to apply
a dynamical framework that can accommodate potentially large momentum-space anisotropies.

In order to take into account plasma momentum-space anisotropy, 
the phase-space distribution of gluons in the local rest frame is assumed to be given 
by~\cite{Dumitru:2007hy,Romatschke:2003ms,Mrowczynski:2004kv,Romatschke:2004jh,Schenke:2006fz}
\begin{equation}
f(t,{\bf x},{\bf p}) = f_{\rm iso}\left(\sqrt{{\bf p}^2+\xi({\bf p}\cdot{\bf
n})^2 }  / p_{\rm hard} \right) ,  \label{eq:f_aniso}
\end{equation}
where $f_{\rm iso}$ is an isotropic distribution which in thermal equilibrium
is given by a Bose-Einstein distribution, $\xi$ is the momentum-space 
anisotropy parameter, and $p_{\rm hard}$ is a momentum scale which specifies the typical momentum
of the particles in the plasma and can be identified with the temperature
in the limit of thermal isotropic ($\xi\!=\!0$) equilibrium.
The two parameters $p_{\rm hard}$ and $\xi$
can, in general, depend on 
proper time and position; however, we do not indicate this explicitly for
compactness of notation.

In general one finds that the heavy quark potential has real and imaginary parts, $V = \Re[V] + i \Im[V]$.  
One can determine the real part of the heavy-quark potential in the non-relativistic limit 
from the Fourier transform of the 00-component of the static gluon propagator.  In an anisotropic plasma with a distribution
function given by Eq.~(\ref{eq:f_aniso}) one finds \cite{Dumitru:2009ni}, at leading order in the strong coupling constant,
\begin{eqnarray}
V({\bf{r}},\xi) &=& -g^2 C_F\int \frac{d^3{\bf{p}}}{(2\pi)^3} \,
e^{i{\bf{p \cdot r}}}\Delta^{00}(\omega=0, \bf{p},\xi) \, , \\
&=& -g^2 C_F\int \frac{d^3{\bf{p}}}{(2\pi)^3} \,
e^{i{\bf{p \cdot r}}} \frac{{\bf{p}}^2+m_\alpha^2+m_\gamma^2}
 {({\bf{p}}^2 + m_\alpha^2 +
     m_\gamma^2)({\bf{p}}^2+m_\beta^2)-m_\delta^4}~, \label{eq:FT_D00}
\end{eqnarray}
where $g$ is the strong coupling constant and $C_F = (N_c^2-1)/(2 N_c)$ is the quadratic Casimir
of the fundamental representation of $SU(N_c)$.  The mass scales $m_\alpha$, $m_\beta$, $m_\gamma$,
and $m_\delta$ are listed in Ref.~\cite{Dumitru:2009ni}.  One can
factorize the denominator of (\ref{eq:FT_D00}) by introducing 
\beq
2 m_{\pm}^2 \equiv M^2 \pm \sqrt{M^4-4(m_\beta^2(m_\alpha^2+m_\gamma^2)-m_\delta^4)} \; ,
\label{mpm}
\eeq
with $ M^2 \equiv m_\alpha^2+m_\beta^2+m_\gamma^2$ \cite{Romatschke:2003ms}.  This
allows us to write
\beq
V({\bf{r}},\xi) = -g^2 C_F\int \frac{d^3{\bf{p}}}{(2\pi)^3} \,
e^{i{\bf{p \cdot r}}} \frac{{\bf{p}}^2+m_\alpha^2+m_\gamma^2}
 {({\bf{p}}^2 + m_+^2)({\bf{p}}^2 + m_-^2)}~. \label{eq:FT_D00_factorized}
 \label{eq:realV}
\eeq
In general one must evaluate (\ref{eq:FT_D00_factorized}) numerically.  The integration can be
reduced to a two-dimensional integral over a polar angle, $\theta$, and the length of the 
three-momentum, $p$.  However, there can be poles in the integration domain due to the 
fact that $m_-^2$ can be negative for certain polar angles and momenta 
\cite{Romatschke:2003ms}.  These poles are first order and can be dealt with using a 
principle-part prescription.  After evaluating the integral numerically, one finds that the resulting 
potential is very well described by a Debye-screened Coulomb potential with an anisotropic Debye
mass $\mu$
\beq
\Re[V(r)] = - \frac{g^2 C_F}{4 \pi r} \, e^{-\mu r}~,  \label{eq:V_iso_shortrange_model}
\eeq
where
\begin{equation}
\left(\frac{\mu}{m_D}\right)^{-4} =  
1 + \xi\left(a - \frac{2^b(a-1)+(1+\xi)^{1/8}}{(3+\xi)^b}\right) 
\left(1 + \frac{c(\theta) (1+\xi)^d}{(1+ e\xi^2)} \right) \, ,
\label{eq:muparam}
\end{equation}
The coefficients a-e are fixed by (a) requiring that the small and large anisotropy limits of the
analytic potential are reproduced and (b) fitting to the numerical results obtained by direct integration of
Eq.~(\ref{eq:realV}) \cite{Strickland:2011aa}.

Here we will focus on a model in which the real part of the potential is obtained from internal energy of the 
system since models based on the free energy seem to be incapable of reproducing either the 
LHC or RHIC  $R_{AA}[\Upsilon]$.  The real part of the potential is given by 
\cite{Strickland:2011aa}
\begin{eqnarray}
\Re[V] =  -\frac{a}{r} \left(1+\mu \, r\right) e^{-\mu \, r }
+ \frac{2\sigma}{\mu}\left[1-e^{-\mu \, r }\right]
- \sigma \,r\, e^{-\mu \, r } -  \frac{0.8\,\sigma}{m_Q^2 r} 
\, , \label{eq:real_pot_model_B}
\end{eqnarray}
where $a=0.385$ and $\sigma = 0.223\;{\rm GeV}^2$ \cite{Petreczky:2010yn} and the
last term is a temperature- and spin-independent finite quark mass correction taken from 
Ref.~\cite{Bali:1997am}.  In this
expression $\mu = {\cal G}(\xi,\theta) m_D$ is an anisotropic Debye mass where ${\cal G}$ 
is a rather complicated function (see Eq.~(\ref{eq:muparam}) above) which depends on the degree of plasma momentum-space anisotropy, 
$\xi$, and the angle of the line connecting the quark-antiquark pair with respect to the beamline 
direction, $\theta$, and  $m_D$ is the isotropic Debye mass \cite{Strickland:2011aa}.  In the limit 
$\xi \rightarrow 0$ one has ${\cal G} = 1$.  In the figures in the results section, results obtained
with Eq.~(\ref{eq:real_pot_model_B}) are often labeled as ``Potential Model B''.

The imaginary part of the potential $\Im[V]$ is obtained from a leading order perturbative calculation 
which was performed in the small anisotropy limit 
\begin{equation} 
\Im[V] = - \alpha_s C_F T \left\{ \phi(r/m_D) - \xi \left[ \psi_1(r/m_D,
\theta)+\psi_2(r/m_D, \theta)\right] \right\} ,
\label{impot}
\end{equation}
where $\phi$, $\psi_1$, and $\psi_2$ can be expressed in terms of hypergeometric functions 
\cite{Dumitru:2009fy}.
After combining the real and imaginary parts of the potential, the 3d Schr\"odinger equation 
is solved numerically to obtain the real and imaginary parts of the binding energy as a function 
of $\xi$ and $p_{\rm hard}$ \cite{Margotta:2011ta,Strickland:2009ft}.  The imaginary part of 
the binding energy is related to the width of the state
\beq
\Gamma(\tau,{\bf x}_\perp,\varsigma) = 
\left\{
\begin{array}{ll}
2 \Im[E_{\rm bind}(\tau,{\bf x}_\perp,\varsigma)]  & \;\;\;\;\; \Re[E_{\rm bind}(\tau,{\bf x}_\perp,\varsigma)] >0 \, , \\
10\;{\rm GeV}  & \;\;\;\;\; \Re[E_{\rm bind}(\tau,{\bf x}_\perp,\varsigma)] \le 0 \, , \\
\end{array}
\right.
\label{eq:width}
\eeq
where $\varsigma = {\rm arctanh}(z/t)$ is the spatial rapidity.  The value of 10 GeV in the 
second case is chosen to be large in order to quickly suppress states which are fully unbound 
(which is the case when the real part of the binding energy is negative).

\subsection{Dynamical Model}

The dynamical model used gives the spatio-temporal evolution of the typical transverse
momentum of the plasma partons, $p_{\rm hard}(\tau,{\bf x})$, and the plasma 
momentum-space anisotropy, $\xi(\tau,{\bf x})$, both of which are specified in the local
rest frame of the plasma.  The widths obtained from solution of the 3d Schr\"odinger equation
are then integrated and exponentiated to compute the relative number of states remaining 
at a given proper time.  This quantity is then averaged over the transverse plane taking
into account the local conditions in the plasma and weighting by the spatial probability 
distribution for bottomonium production which is given by the number of binary collisions
computed in the Glauber model with a Woods-Saxon distribution for each nucleus.  For the
temporal integration the initial time is set by the formation time of the state in question.
The resulting $R_{AA}$ is a function of the transverse momentum, $p_T$, the rapidity
$\varsigma$, and the nuclear impact parameter $b$.  To compare to experimental results
transverse momentum cuts are applied assuming a $n(p_T) = n_0 E_T^{-4}$ spectrum.  
In addition, any cuts on the rapidity due to detector acceptance and centrality are applied
as needed.  For details of the dynamical model and $R_{AA}$ computation we refer the 
reader to Ref.~\cite{Strickland:2011aa}.

\begin{figure}[t]
\includegraphics[width=0.49\textwidth]{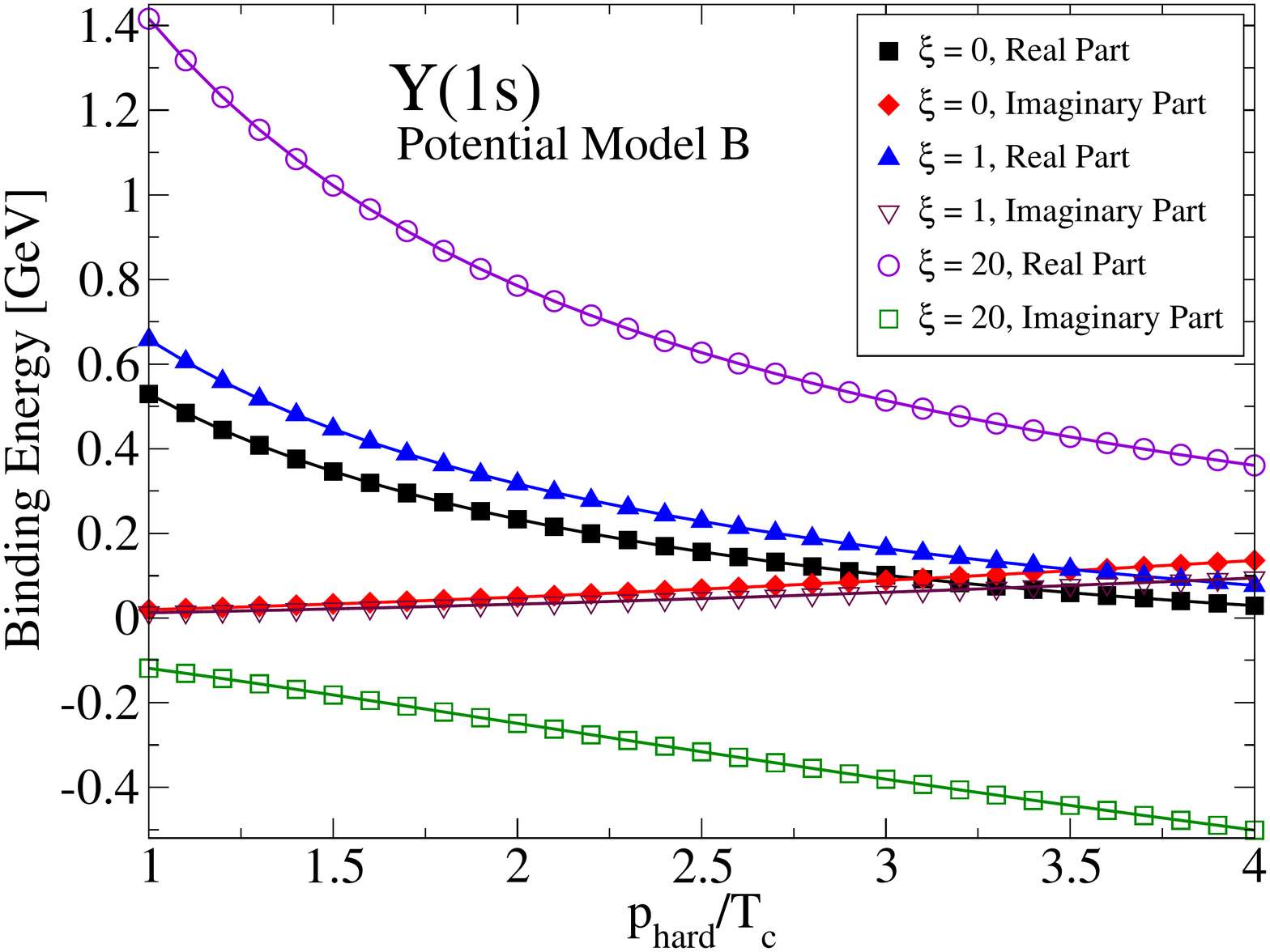}
\hspace{2mm}
\includegraphics[width=0.49\textwidth]{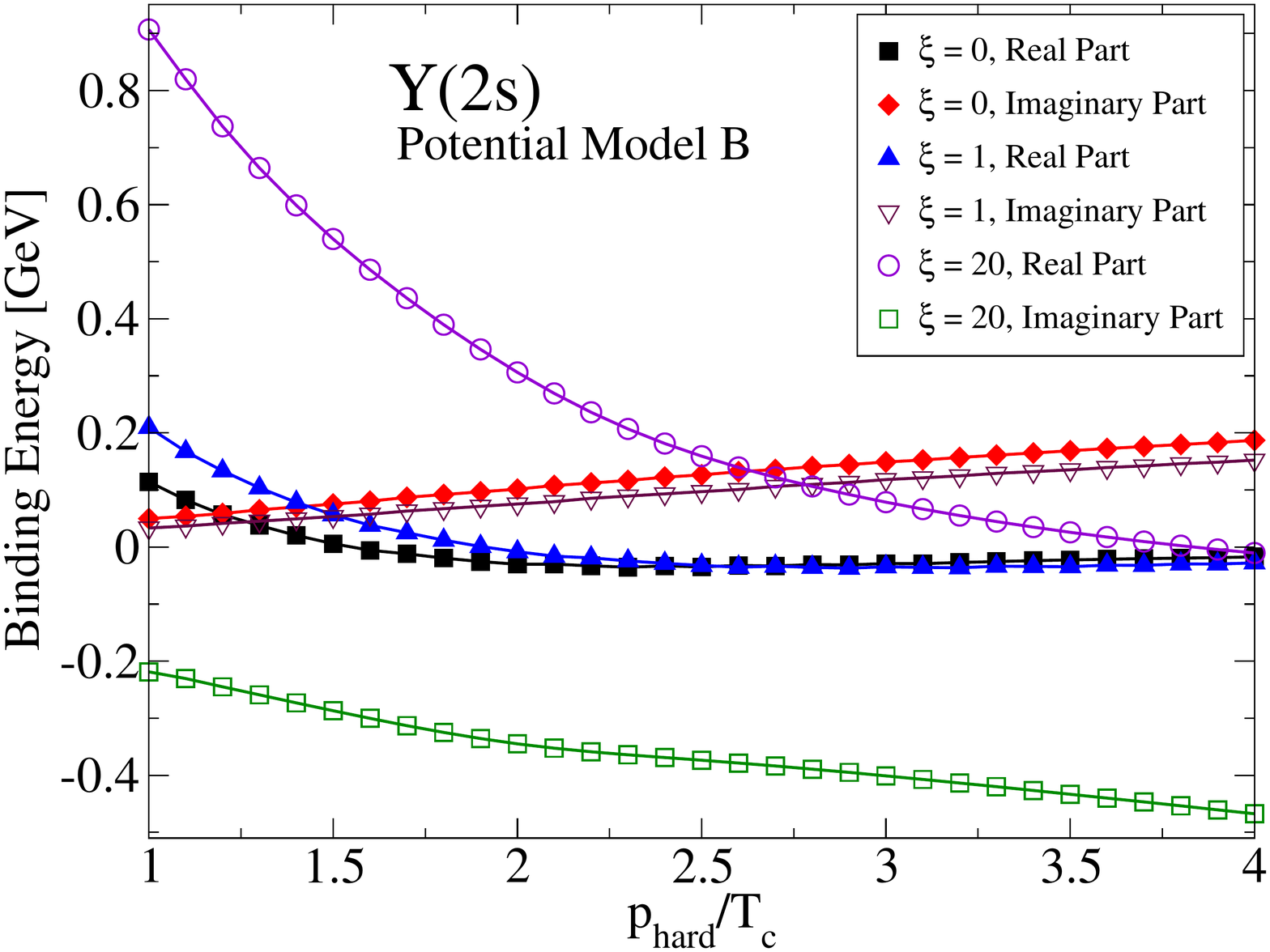}
\caption{
Real and imaginary parts of the $\Upsilon(1s)$ (left) and $\Upsilon(2s)$ (right)  binding energies 
as a function of the hard momentum scale, $p_{\rm hard}$ for different levels of momentum-space
anisotropy $\xi$. 
}
\label{fig:binding}
\end{figure}

\subsection{Initial conditions}

For the initial conditions we use a Woods-Saxon distribution for each nucleus and determine 
the transverse dependence of the initial temperature via the third root of the number of 
participants (wounded nucleons). In the spatial rapidity direction we have investigated two possible
temperature profiles:  (a) a broad plateau containing a boost-invariant central region with 
Gaussian limited-fragmentation at large rapidity \cite{Strickland:2012cq}
\beq
 n(\varsigma) = n_0 \exp(-(|\varsigma|-\varsigma_{\rm flat}/2)^2/2 \sigma_\varsigma^2)
\; \Theta(|\varsigma|-\varsigma_{\rm flat}/2) \, ,
\label{eq:rapplateau}
 \eeq
where $\varsigma_{\rm flat}=10$ is the width of the central rapidity plateau,
$\sigma_\varsigma = 0.5$ is the width of the limited fragmentation tails,
and $n_0$ is the number density at central rapidity \cite{Schenke:2011tv}; and (b) a 
Gaussian motivated by low-energy fits to pion spectra
\begin{equation}
\label{eq:yprofile}
 n(\varsigma) = n_0 \exp(-\varsigma^2/2\sigma_\varsigma^2) 
\quad
{\rm with}
\quad
 \sigma_\varsigma^2=0.64 \cdot 8 \, c_s^2 \ln \left(\sqrt{s_{NN}}/2 m_p\right) /3(1-c_s^4)
 \; ,
 \label{eq:rapgauss}
\end{equation}
where $c_s = 1/\sqrt{3}$ is the sound velocity, $m_p = 0.938$ GeV is the proton mass, 
and $\sqrt{s_{NN}}$ is the nucleon-nucleon center-of-mass energy 
\cite{Strickland:2011aa,Bleicher:2005tb}.
The temperature distribution is given by $T \sim n^{1/3}$.
We note that Eq.~(\ref{eq:rapplateau})  has the advantage that it has been tuned to successfully
describe the rapidity dependence of the elliptic flow in LHC heavy ion collisions.

\section{Results and Conclusions}

In Fig.~\ref{fig:binding} we show the real and imaginary parts of the $\Upsilon(1s)$ and $\Upsilon(2s)$
binding energies as a function of the hard momentum scale, $p_{\rm hard}$, for $\xi \in \{0,1,20\}$.  
Defining the disassociation scale as the value of $p_{\rm hard}$ at which the real and imaginary parts of the 
binding energy become equal, one finds isotropic dissociation scales of 593 MeV for the $\Upsilon(1s)$ 
and 228 MeV for the $\Upsilon(2s)$.  In addition, as one can see from this figure, as the level of momentum
space anisotropy increases, one finds that the disassociation scale increases.  This means that in regions of
the plasma where there is a high degree of momentum-space anisotropy, the states will be less suppressed.

\begin{figure}[t]
\includegraphics[width=0.49\textwidth]{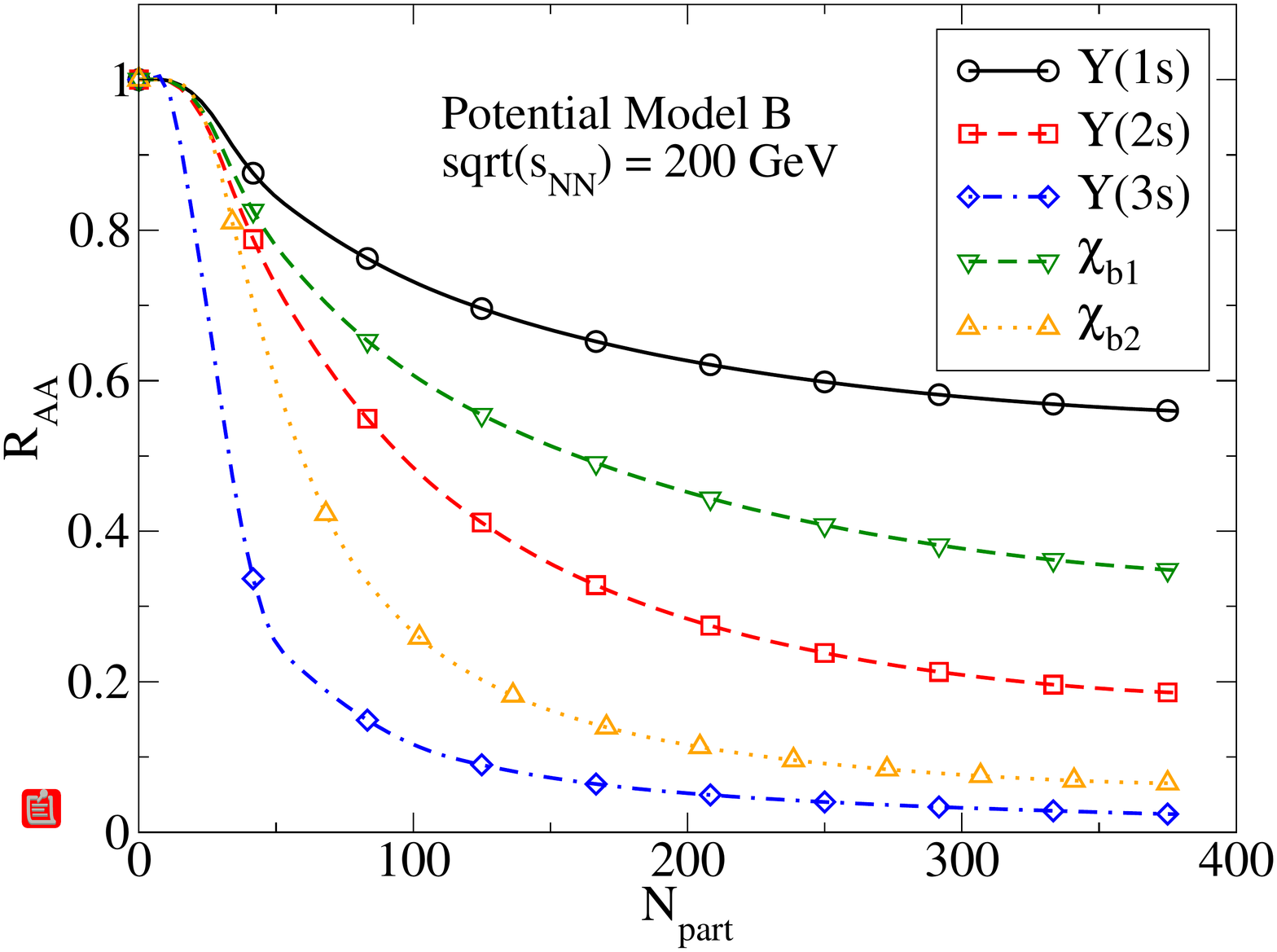}
\includegraphics[width=0.49\textwidth]{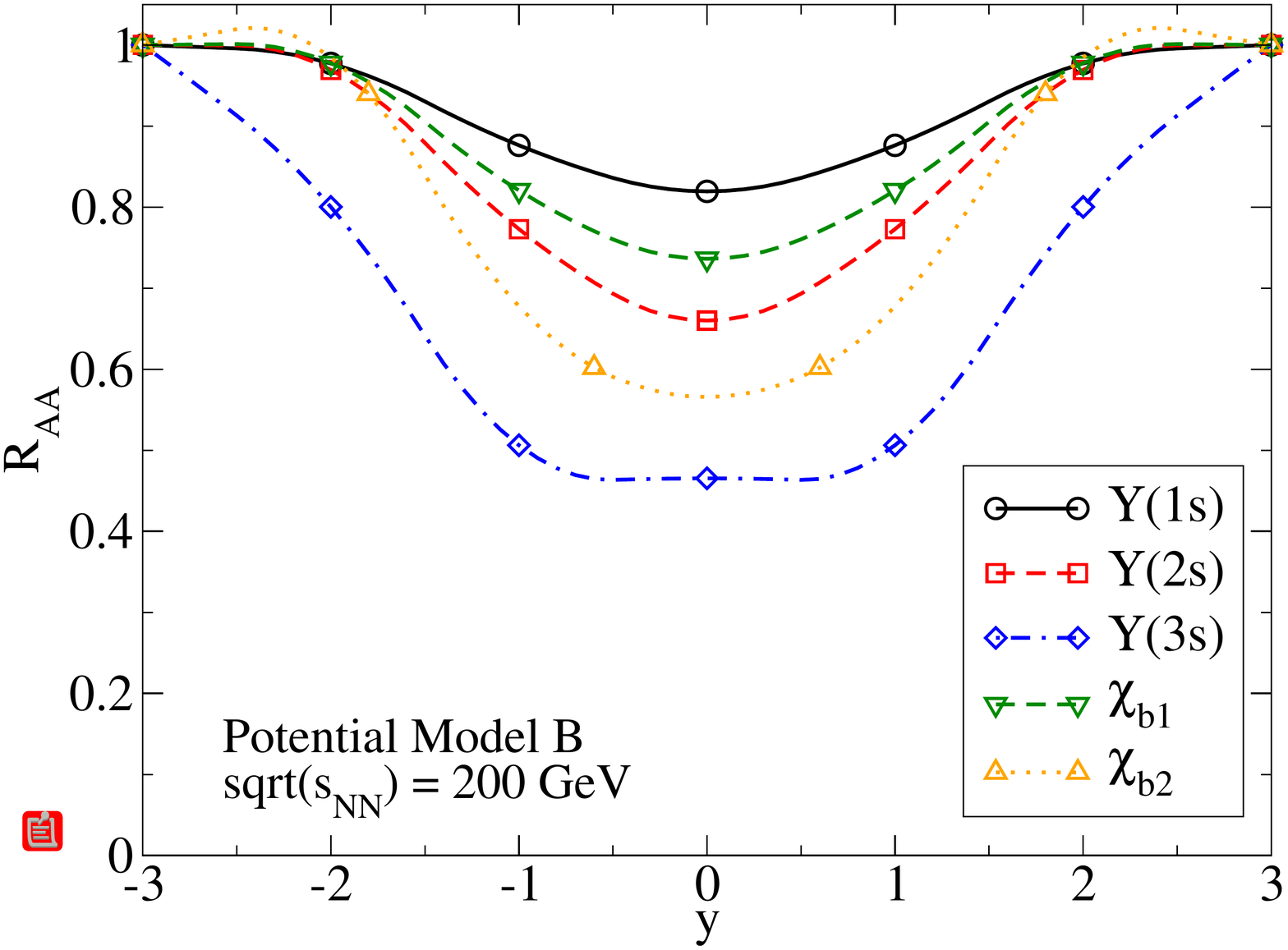}
\\
\includegraphics[width=0.49\textwidth]{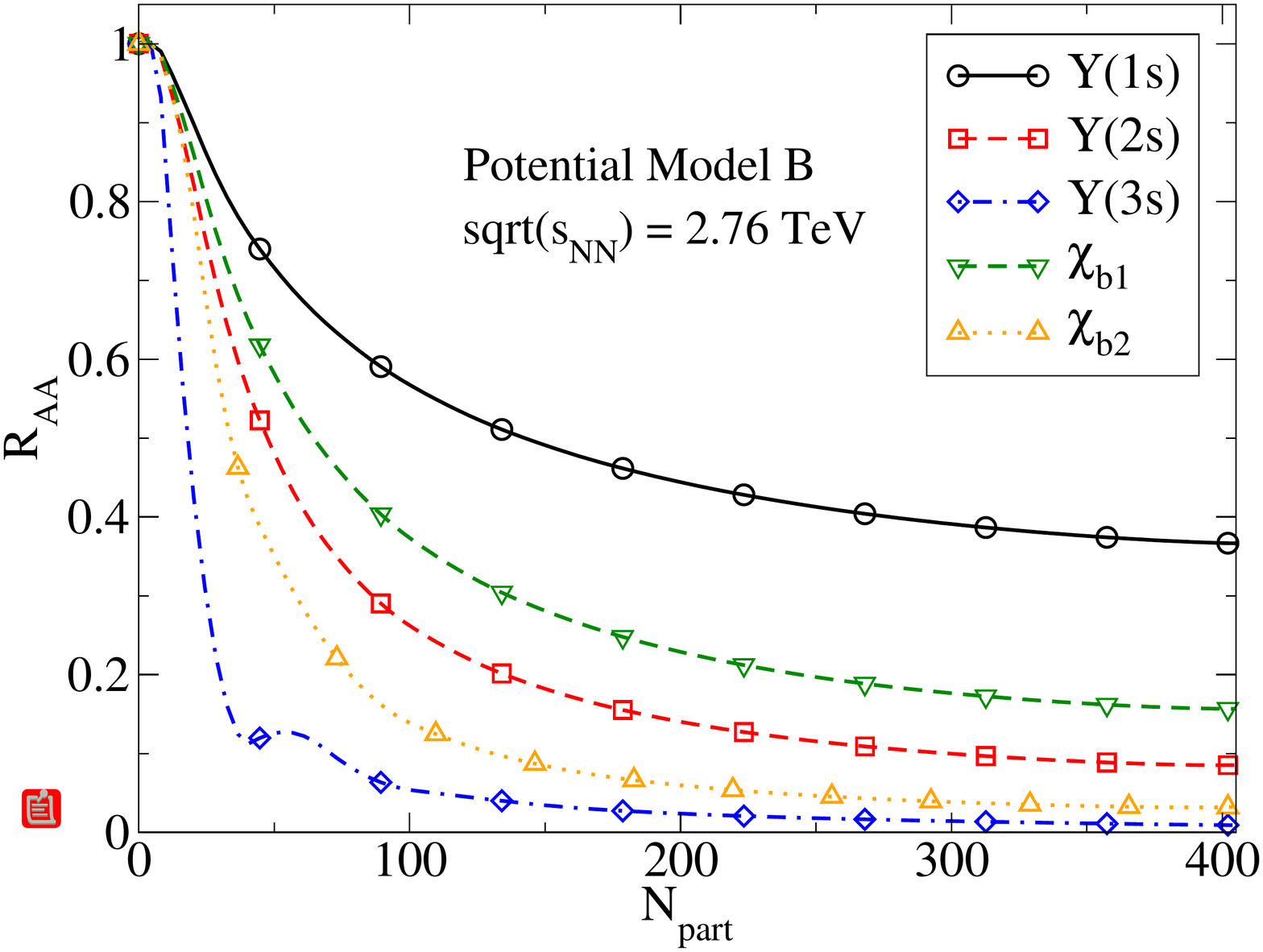}
\includegraphics[width=0.49\textwidth]{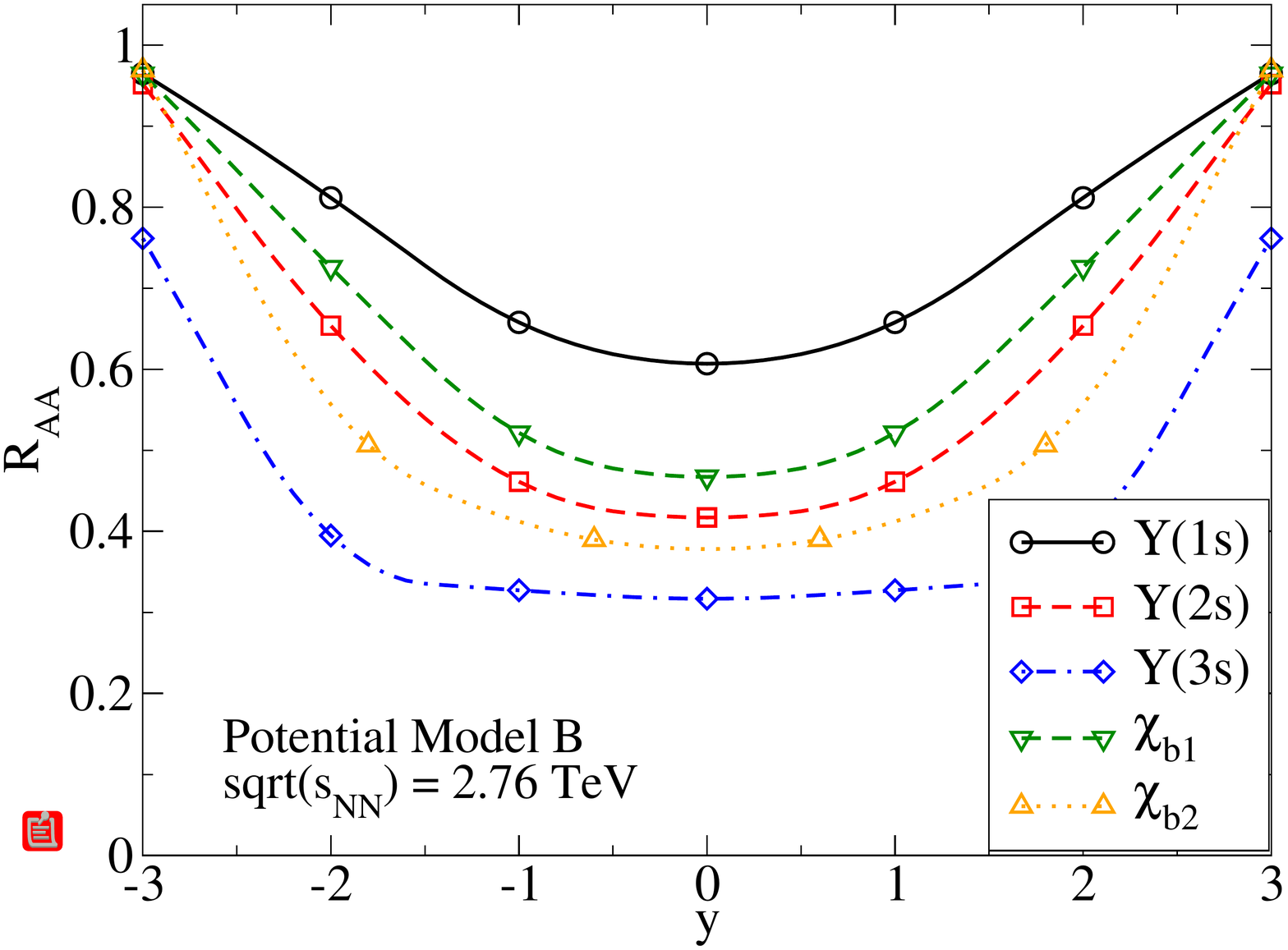}
\caption{
RHIC and LHC suppression factor $R_{AA}$ for the $\Upsilon(1s)$, $\Upsilon(2s)$, $\Upsilon(3s)$, $\chi_{b1}$, and $\chi_{b2}$ 
states as a function of the number of participants (left) and rapidity (right).  In the top plots we used 
$\sqrt{s_{NN}} = 200$ GeV and in the bottom plots we used $\sqrt{s_{NN}} = 2.76$ TeV.  In both cases, we assumed a shear viscosity 
to entropy density ratio of $4 \pi \eta/{\cal S} = 1$, and 
implemented cuts of $0 < p_T < 20$ GeV and (left) rapidity $|y| < 0.5$ for RHIC and $|y| < 2.4$ for LHC (right) centrality 0-100\%.
}
\label{fig:raa-potb}
\end{figure}

In Fig.~\ref{fig:raa-potb} we show the predicted suppression factor $R_{AA}$ for the $\Upsilon(1s)$, 
$\Upsilon(2s)$, $\Upsilon(3s)$, $\chi_{b1}$, and $\chi_{b2}$ states as a function of the number of participants (left) and rapidity (right).
In this figure we see clear signs of sequential
suppression, with the higher excited states having stronger suppression than the ground state.  However, we note
that even for states that are ``melted'' at relatively low central temperatures, we still obtain a non-vanishing suppression
factor for these states.  This is due to the fact that near the edges, where the temperature is lower, one does not
see suppression of the states.  Upon performing the geometrical average, we see
that a large fraction of the states produced can survive even when the central temperature of the plasma is above 
their naive dissociation temperature.

\begin{figure}[t]
\includegraphics[width=0.52\textwidth]{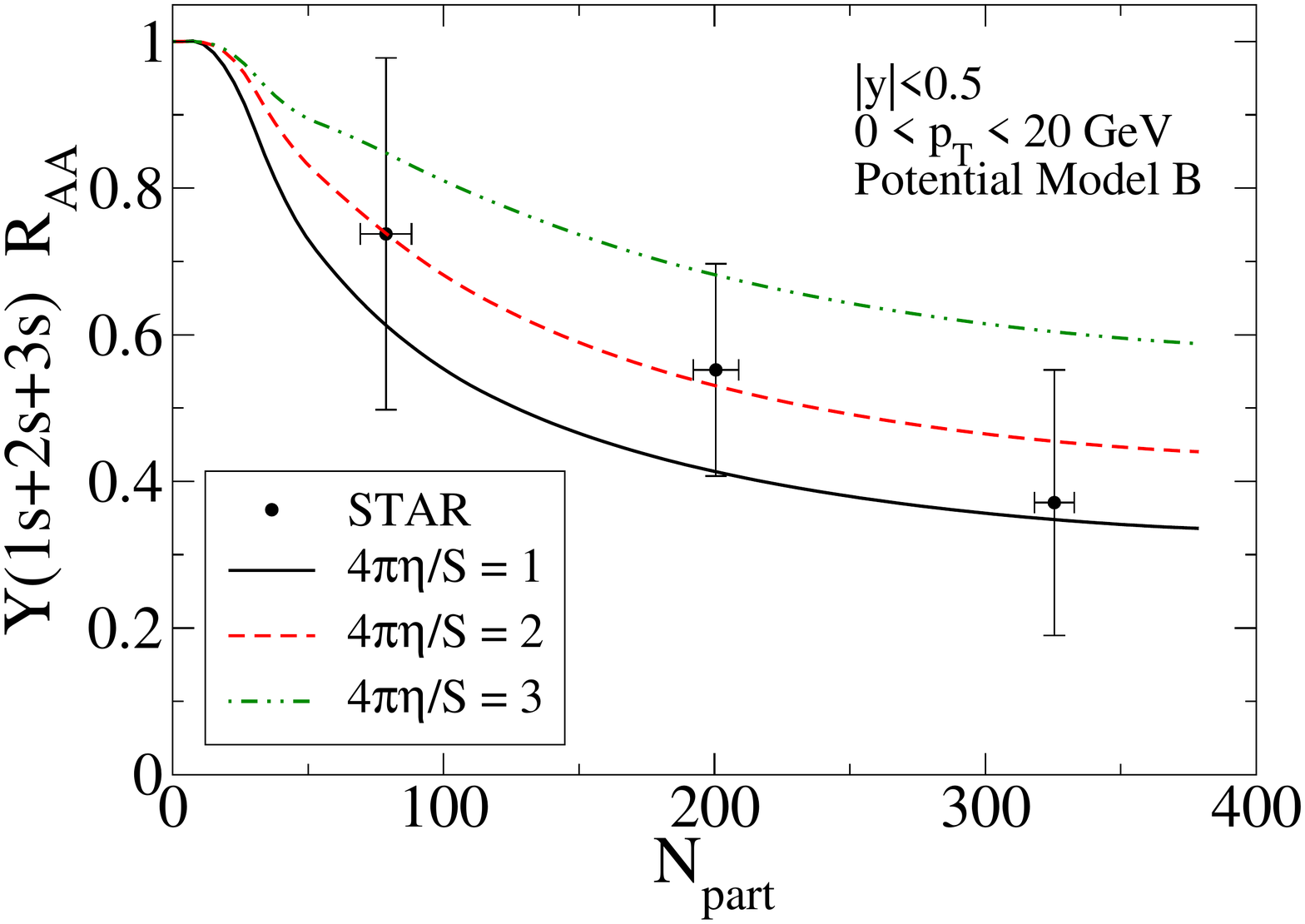}
\includegraphics[width=0.47\textwidth]{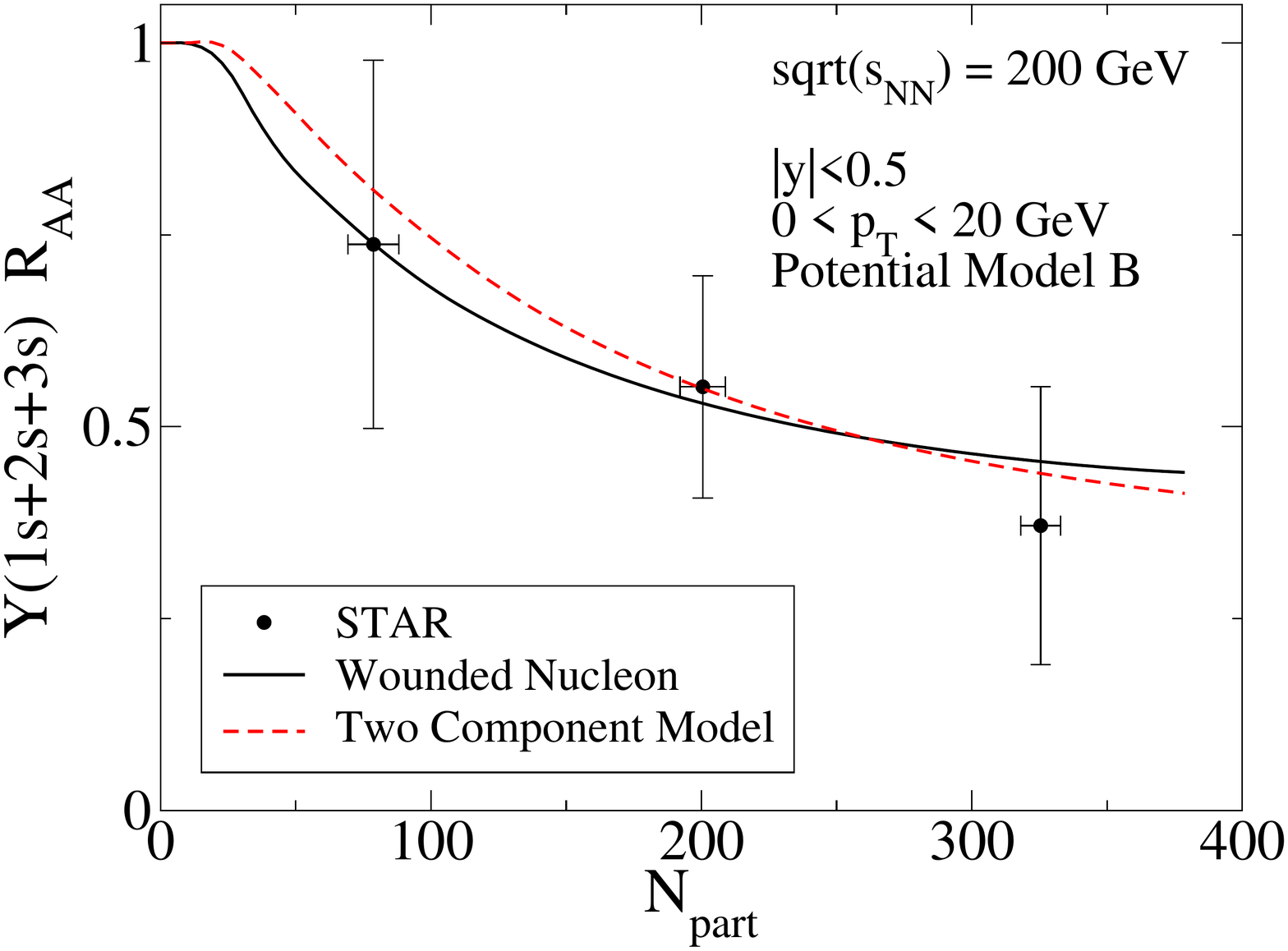}
\caption{
RHIC $\Upsilon(1s+2s+3s)$ suppression factor (left) compared with 
experimental data from the STAR Collaboration \cite{Reed:2011fr}.  The
three different lines correspond to different assumptions for the shear viscosity to entropy ratio 
$4 \pi \eta/{\cal S} \in \{1,2,3\}$.  In both plots we used $\sqrt{s_{NN}} = 200$ GeV and 
implemented cuts of $0 < p_T < 20$ GeV and $|y| < 0.5$.  On the right, the solid black line is the 
result obtained assuming wounded nucleon initial conditions and the dashed red line is the result obtained used 
a two component model with $\alpha=0.145$.
}
\label{fig:y1s2s3s-rhic-potb}
\end{figure}

In Fig.~\ref{fig:y1s2s3s-rhic-potb} (left) we plot $R_{AA}[\Upsilon(1s+2s+3s)] $ and compare with 
experimental data from the STAR Collaboration \cite{Reed:2011fr}.
As can be seen from this figure, the model does a reasonably good job
of reproducing the existing STAR data for $R_{AA}[\Upsilon(1s+2s+3s)]$.  From the left
panel we can obtain an estimate of $\eta/{\cal S}$: $0.08 < \eta/{\cal S} < 0.24$. 
In Fig.~\ref{fig:y1s2s3s-rhic-potb} (right) we show the results obtained
for $R_{AA}[\Upsilon(1s+2s+3s)]$ at RHIC energies for two different types of initial conditions (Glauber
and mixed Glauber plus binary collision scaling).  For both lines shown in Fig.~\ref{fig:y1s2s3s-rhic-potb} (right) 
we have assumed $4 \pi \eta/S = 2$.
Because changing the initial condition type affects particle multiplicities we have adjusted the initial temperature
at RHIC energies from 433 MeV to 461 MeV in order to keep the charged particle 
multiplicity fixed at $dN_{ch}/dy = 620$.  As can be seen from Fig.~\ref{fig:y1s2s3s-rhic-potb} (right),
for peripheral collisions there is a larger dependence on the choice of initial condition type, while for central collisions
the result obtained is not much affected by the choice of initial condition.  This is to be contrasted with the dependence
on the assumed value of $\eta/{\cal S}$ which affects the suppression at all centralities.

\begin{figure}[t]
\includegraphics[width=0.49\textwidth]{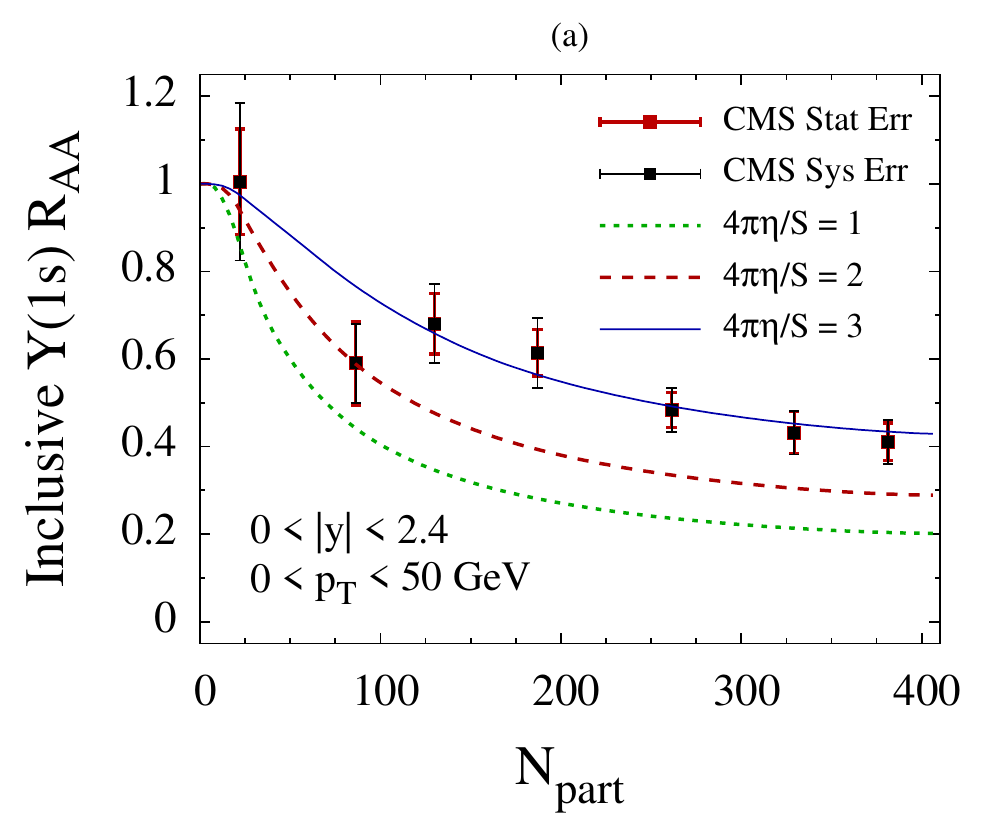}
\hspace{2mm}
\includegraphics[width=0.49\textwidth]{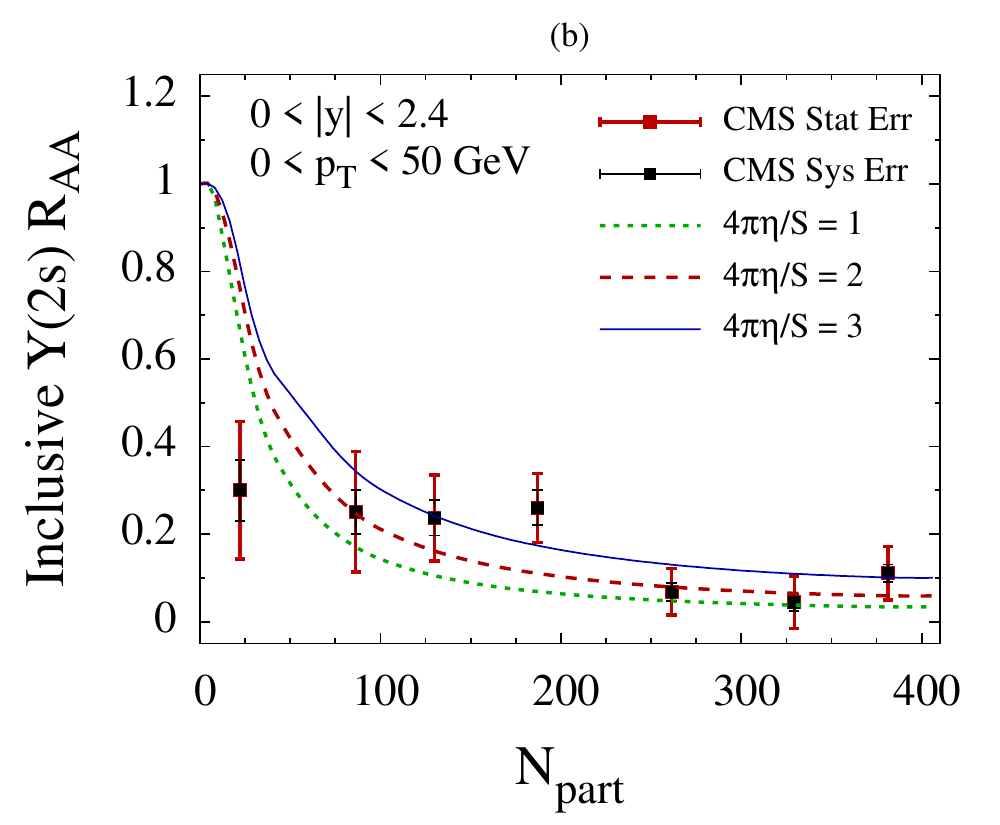}
\caption{Predictions for the central rapidity inclusive (a) $\Upsilon(1s)$ and (b) 
$\Upsilon(2s)$ suppression including feed-down as a function of $N_{\rm part}$ along with 
recent data from the CMS collaboration.  The
three different lines correspond to different assumptions for the shear viscosity to entropy ratio 
$4 \pi \eta/{\cal S} \in \{1,2,3\}$. 
}
\label{fig:cmscompare}
\end{figure}

\begin{figure}[t]
\includegraphics[width=0.49\textwidth]{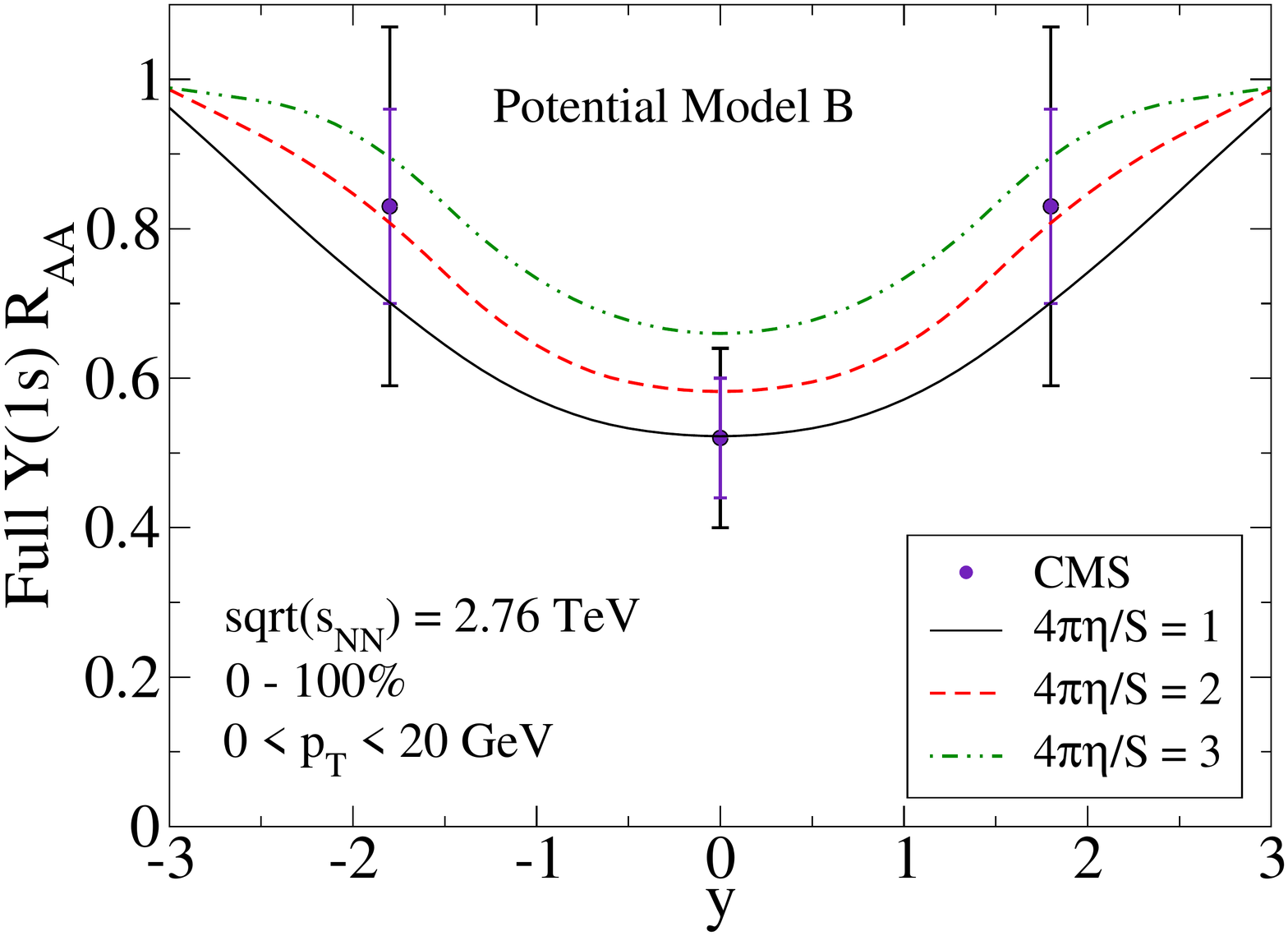}
\hspace{2mm}
\includegraphics[width=0.49\textwidth]{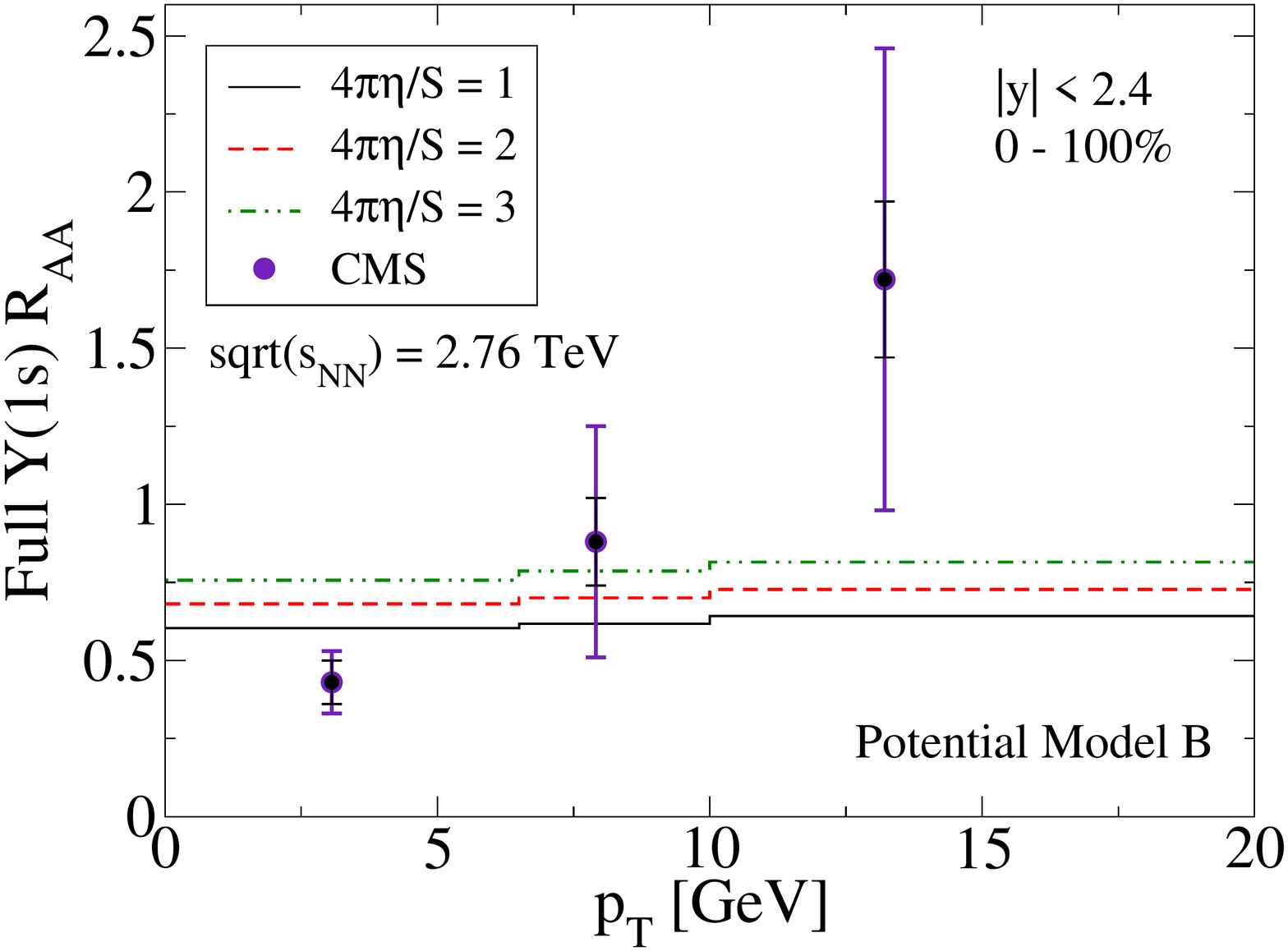}
\caption{
LHC inclusive suppression factor $R_{AA}$ for the $\Upsilon(1s)$ including feed down effects 
as a function of rapidity and transverse momentum compared to experimental data are from the CMS 
Collaboration \cite{HIN-10-006}. 
}
\label{fig:y1s-lhc}
\end{figure}

When considering the suppression of the $\Upsilon(1s)$ and $\Upsilon(2s)$ states it is 
important to include the effect of feed-down from higher excited states.  In pp collisions 
only approximately 51\% of $\Upsilon(1s)$ states come from direct production and similarly for 
the $\Upsilon(2s)$.  One can compute the inclusive suppression of a state using 
$R_{AA}^{\rm full}[\Upsilon(ns)] = \sum_{i\,\in\,{\rm states}} f_i \,R_{i,AA}$
where $f_i$ are the feed-down fractions and $R_{i,AA}$ is the direct suppression
of each state which decays into the $\Upsilon(ns)$ state being considered.
Here we will use $f_i = \{0.510,0.107,0.008,0.27,0.105\}$ for the $\Upsilon(1s)$, $
\Upsilon(2s)$, $\Upsilon(3s)$,  $\chi_{b1}$, and $\chi_{b2}$ feed-down to $\Upsilon(1s)$, 
respectively \cite{Affolder:1999wm}. For the inclusive $\Upsilon(2s)$ production we use 
$f_i = \{0.500,0.500\}$ for the  $\Upsilon(2s)$ and $\Upsilon(3s)$ states, respectively.
For details of the computation of the direct $R_{AA}$ for each state see 
Ref.~\cite{Strickland:2011aa}.

In Fig.~\ref{fig:cmscompare} we compare to recent data on the inclusive $\Upsilon(1s)$ and $\Upsilon(2s)$ 
suppression available from the CMS collaboration 
\cite{Chatrchyan:2012fr}.  For this figure we
used a broad rapidity plateau as the initial density profile as specified in 
Eq.~(\ref{eq:rapplateau}).  The central temperatures 
were taken to be $T_0 = \{520,504,494\}~$MeV at $\tau_0 = 0.3$~fm/c 
for $4 \pi \eta/S = \{1,2,3\}$, respectively, in 
order to fix the final charged multiplicity to $dN_{\rm ch}/dy = 1400$ in each case.  As 
can been seen from this figure, the predictions agree reasonably well with the available
data.  The data seem to prefer the largest value of $\eta/S$ shown; however, there is a
$\pm 14\%$, $\pm 21\%$ 1s, 2s global uncertainty reported by CMS, making
it hard to draw firm conclusions.

In Fig.~\ref{fig:y1s-lhc} we show the inclusive suppression factor $R_{AA}^{\rm full}[\Upsilon(1s)]$. 
Comparing to the available CMS data \cite{HIN-10-006}  we can obtain an estimate for
$\eta/{\cal S}$ at LHC energies:  $0.08 < \eta/{\cal S} < 0.24$ which is the same range obtained from the STAR
data obtained with gold-gold collisions at lower energies.  As before, more precisely determining $\eta/{\cal S}$ 
will require more data from the LHC which should be forthcoming in the near future.

\section{Conclusions}

In this proceedings contribution we have reviewed recent calculations of 
bottomonium suppression in RHIC and LHC heavy ion collisions.  We presented comparisons of prior 
predictions with data from the CMS Collaboration for inclusive $\Upsilon(1s)$ and $\Upsilon(2s)$ 
suppression and data from the STAR Collaboration for inclusive $\Upsilon(1s+2s+3s)$ suppression.
The underlying calculations employed a complex-valued potential which incorporates both screening
and in-medium dissociation of the states under consideration.  We solved for the resulting real and
imaginary parts of the binding energy of the $\Upsilon(1s)$, $\Upsilon(2s)$, $\Upsilon(3s)$, $\chi_{b1}$, 
and $\chi_{b2}$ states.  We then folded this information together with the real-time evolution 
of both the typical momentum of the plasma particles ($p_{\rm hard}$) and their momentum-space
anisotropy ($\xi$).  We demonstrated that the resulting inclusive suppression of $\Upsilon$ states is
in good agreement with available data.  We note in closing that there are now very interesting 
developments concerning quarkonium in an anisotropic medium obtained employing the conjectured 
AdS/CFT correspondence \cite{Giataganas:2012zy,Rebhan:2012bw,Chernicoff:2012bu}.  These methods 
offer some hope to determine the temperature and momentum-space anisotropy dependence of the long range 
part of the potential from first principles, albeit in a theory that is not QCD.


\section*{Acknowledgments}

This work was supported by NSF grant No. PHY-1068765 and the Helmholtz International Center 
for FAIR LOEWE program.

\section*{References}

\bibliographystyle{iopart-num}
\bibliography{strickland}

\end{document}